\documentclass[twocolumn,dvipsnames]{aastex62}
\pdfoutput=1 
\usepackage{amsmath,amstext}
\usepackage{apjfonts}
\usepackage[figure,figure*]{hypcap}
\usepackage{ulem}
\usepackage{xspace}

\usepackage{graphicx}
\usepackage{natbib}
\usepackage{amssymb}
\usepackage{amsmath}
\usepackage{array,booktabs}
\usepackage{mathptmx}
\usepackage{booktabs}
\usepackage{hyperref}
\usepackage{float}

\newcommand{\msun}{\,{M_{\odot}}}

\newcommand{\mbh}{\,{M_{\rm BH}}}

\newcommand{\Egiso}{\,{E_{\gamma,{\rm iso}}}}
\newcommand{\Egisomin}{\,{E_{\gamma,{\rm iso,min}}}}
\newcommand{\Egisomax}{\,{E_{\gamma,{\rm iso,max}}}}
\usepackage{float}
\usepackage{newfloat}
\DeclareFloatingEnvironment[name={Extended Data Figure}]{extdatafig}
\usepackage{lineno}

\shorttitle{Jetted and Turbulent Stellar Deaths: New GW Sources}
\shortauthors{Gottlieb et al.}

\begin{document}

\title{Jetted and Turbulent Stellar Deaths: New LVK-Detectable Gravitational Wave Sources} 

	\author[0000-0003-3115-2456]{Ore Gottlieb}
	\email{ore@northwestern.edu}
	\affiliation{Center for Interdisciplinary Exploration \& Research in Astrophysics (CIERA), Physics \& Astronomy, Northwestern University, Evanston, IL 60202, USA}

	\author[0000-0002-7205-6367]{Hiroki Nagakura}
	\affiliation{Division of Science, National Astronomical Observatory of Japan, 2-21-1 Osawa, Mitaka, Tokyo 181-8588, Japan}

	\author[0000-0002-9182-2047]{Alexander Tchekhovskoy}
	\affiliation{Center for Interdisciplinary Exploration \& Research in Astrophysics (CIERA), Physics \& Astronomy, Northwestern University, Evanston, IL 60202, USA}
    
	\author[0000-0002-5554-8896]{Priyamvada Natarajan}
	\affiliation{Department of Astronomy, Yale University, 52 Hillhouse Avenue, New Haven, CT 06520, USA}
    \affiliation{Department of Physics, Yale University, P.O. Box 208121, New Haven, CT 06520, USA}
    \affiliation{Black Hole Initiative, Harvard University, 20 Garden Street, Cambridge MA 02138, USA}

	\author[0000-0003-2558-3102]{Enrico Ramirez-Ruiz}
	\affiliation{Department of Astronomy and Astrophysics, University of California, Santa Cruz, CA 95064, USA}

    \author[0000-0001-7852-7484]{Sharan Banagiri}
	\affiliation{Center for Interdisciplinary Exploration \& Research in Astrophysics (CIERA), Physics \& Astronomy, Northwestern University, Evanston, IL 60202, USA}

    \author[0000-0003-2982-0005]{Jonatan Jacquemin-Ide}
	\affiliation{Center for Interdisciplinary Exploration \& Research in Astrophysics (CIERA), Physics \& Astronomy, Northwestern University, Evanston, IL 60202, USA}

    \author[0000-0002-5375-8232]{Nick Kaaz}
	\affiliation{Center for Interdisciplinary Exploration \& Research in Astrophysics (CIERA), Physics \& Astronomy, Northwestern University, Evanston, IL 60202, USA}

    \author[0000-0001-9236-5469]{Vicky Kalogera}
	\affiliation{Center for Interdisciplinary Exploration \& Research in Astrophysics (CIERA), Physics \& Astronomy, Northwestern University, Evanston, IL 60202, USA}

\begin{abstract}
      
      Upcoming LIGO/Virgo/KAGRA (LVK) observing runs are expected to detect a variety of inspiralling gravitational-wave (GW) events, that come from black-hole and neutron-star binary mergers. Detection of non-inspiral GW sources is also anticipated. We report the discovery of a new class of non-inspiral GW sources - the end states of massive stars - that can produce the brightest simulated stochastic GW burst signal in LVK bands known to date, and could be detectable in the LVK run A+. Some dying massive stars launch bipolar relativistic jets, which inflate a turbulent energetic bubble -- cocoon -- inside of the star. We simulate such a system using state-of-the-art 3D general-relativistic magnetohydrodynamic simulations and show that these cocoons emit quasi-isotropic GW emission in the LVK band, $\sim 10-100$~Hz, over a characteristic jet activity timescale, $\sim 10-100$~s. Our first-principles simulations show that jets exhibit a wobbling behavior, in which case cocoon-powered GWs might be detected already in LVK run A+, but it is more likely that these GWs will be detected by the third generation GW detectors with estimated rate of $ \sim 10 $ events/year. The detection rate drops to $ \sim 1\% $ of that value if all jets were to feature a traditional axisymmetric structure instead of a wobble. Accompanied by electromagnetic emission from the energetic core-collapse supernova and the cocoon, we predict that collapsars are powerful multi-messenger events.
      
\end{abstract}

\section{Introduction}\label{sec:intro}

Core-collapse supernovae (CCSNe) provide a unique opportunity to study the last stages of stellar life-cycles, the synthesis of heavy elements, and the birth of compact objects \citep{Burrows1986,Woosley2002,Woosley2005}. However, the presence of intervening opaque stellar gas limits the prospects for learning about the underlying physics of the explosion mechanism and the compact object environment from electromagnetic signals alone. Fortuitously, CCSNe produce two extra messengers: neutrinos and gravitational-waves (GWs), both of which carry information from the stellar core to the observer with negligible interference along the way \citep{Bethe1990,Janka1996,Janka2012}. Numerical studies \citep{Ott2009,Kotake2013} have shown that CCSNe can be highly asymmetric, giving rise to a substantial time-dependent gravitational quadrupole moment (and higher modes), which generates GW emission. However, with a small fraction ($ E_{\rm GW} \sim 10^{46}~{\rm erg} $) of the CCSN energy emerging as GWs, the ground-bound based interferometers LIGO/Virgo/KAGRA \citep[LVK;][]{LIGO} and cosmic explorer (CE) can detect only nearby ($ \lesssim 100 $ kpc) events at design sensitivity \citep[e.g.,][]{Srivastava2019}.

A special class of CCSNe -- collapsars \citep{Woosley1993} is associated with long-duration gamma-ray bursts (LGRBs), which originate in energetic jets powered by a rapidly rotating newly-formed compact object, a black-hole (BH) or neutron-star. Their enormous power makes LGRB jets interesting GW sources. The GW frequency is inversely proportional to the timescale over which the metric is perturbed, and for jets, this timescale is set by the longer of the launching and acceleration timescales \citep{Sago2004,Hiramatsu2005,Akiba2013,Birnholtz2013,Yu2020,Leiderschneider2021,Urrutia2022}. Thus, the characteristic duration of LGRBs $ \gtrsim 10 $~s places the GW emission from LGRB jets at the sub-Hz frequency band, too low for LVK, but potentially detectable by the proposed space-based Decihertz Interferometer GW Observatory \citep{Kawamura2011}.
	
\begin{figure}
	\centering
		\includegraphics[scale=0.4]{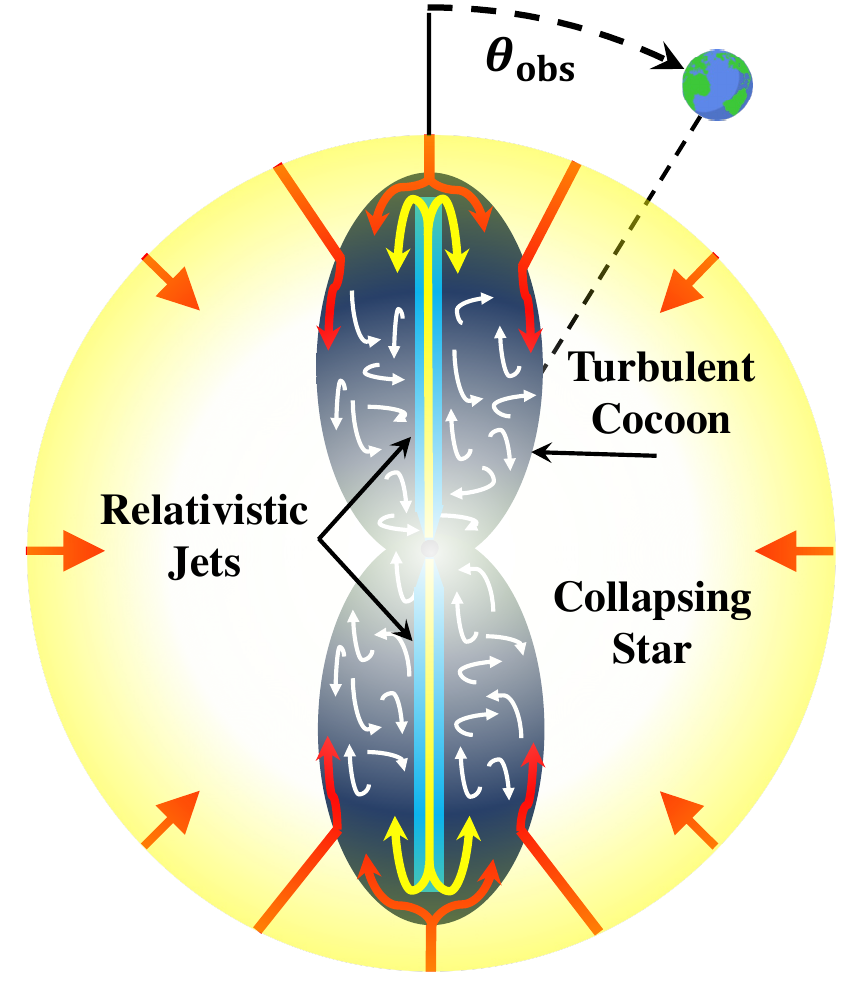}
		\caption{Jets (light blue) inflate the hourglass-shaped cocoon (dark blue) inside of a dying massive star (pale yellow). The jets run into and shock against the collapsing star, forming backflows (yellow arrows). In-falling star runs into and shocks against the backflows (red arrows). The shocked jets and shocked stellar components form the cocoon and turbulently mix inside of it (white arrows). 
		}
		\label{fig:sketch}
	\end{figure}

As the jets drill their way out of the collapsing star, they shock the dense stellar material and build a cocoon -- a hot and turbulent hourglass-shape structure that envelops the jets \citep[Fig.~\ref{fig:sketch};][]{Meszaros2001,RamirezRuiz2002,Matzner2003,Lazzati2005}. The cocoon is generated as long as parts of the jets are moving sub-relativistically inside the star. LGRB jets break out from the star after $ t_b \sim 10 $~s \citep{MacFadyen1999,Zhang2004,Bromberg2016} and spend a comparable amount of time outside of the star before their engine turns off \citep{Bromberg2011}. After breakout, jets are expected to stop depositing energy into the cocoon \citep{RamirezRuiz2002,Lazzati2005,Bromberg2011,Gottlieb2021a}, unless they are intermittent or wobbly \citep{Gottlieb2020b,Gottlieb2021b,Gottlieb2022b}. This implies that the cocoon energy $ E_c \simeq E_j \sim 10^{51} - 10^{52} $~erg \citep{Nakar2017} is similar to the jet energy if the jet propagates along a fixed axis. But $ E_c $ can be even larger than the \emph{observed} jet energy if the jet wobbles. Our simulation indicates that in such a scenario, the cocoon energy could be larger by an order of magnitude, thereby impacting the GW detection rate. Consequently, the detection of cocoon-powered GWs can serve as a valuable tool for investigating the jet structure.

In this {\it Letter}, we show that the cocoon is a promising new GW source for present-day detectors. The cocoon evolves over shorter timescales than the jets, and, as we show, its GW signal lies within the LVK frequency band. 
We first calculate the GW emission using an analytic approximation and then compute it numerically (\S\ref{sec:gw}), and then turn to estimate the detectability of such events (\S\ref{sec:detectability}). We discuss the multi-messenger detection prospects in \S\ref{sec:multimessenger} and conclude in \S\ref{sec:conclusion}.

\section{Gravitational-waves from cocoons}\label{sec:gw}

To calculate the GW emission, we carry out a high resolution 3D general-relativistic magnetohydrodynamic (GRMHD) simulation of LGRB jet formation, propagation and escape from a collapsing star. The simulation is similar to those discussed in \citet{Gottlieb2022b}, but here we double the stellar mass for consistency with LGRB progenitor stars \citep{Kuncarayakti2013}. Due to the scale-free nature of the simulation, doubling the mass results in an equivalent doubling of the energy, while leaving the magnetohydrodynamic evolution of the system unaffected. We also increase the data output frequency, as needed to resolve the GW signal at high frequencies. Overall we post-process more than a petabyte of simulation data output at a high cadence ($ \sim 50 $ kHz sampling frequency). This simulation follows the jet and cocoon from the BH for $ \simeq 3.2 $~s until they reach distance $ \simeq 1.5R_\star $ with energy $ E = E_j + E_c \simeq 10^{53} $ erg, where $ E_c \simeq 2E_j $ and $ R_\star $ is the stellar radius. The simulation details are provided in Appendix \ref{sec:gw_calculation}.

We first provide an order of magnitude estimate of the GW strain, using the familiar quadrupole approximation:
    \begin{equation}\label{eq:h}
	    h \approx \frac{2G}{Dc^4}\frac{d^2Q}{dt^2}\approx \frac{4G}{Dc^4}E\epsilon \simeq 10^{-23}~\frac{100~{\rm Mpc}}{D}\frac{E\epsilon}{10^{52}~{\rm erg}}~,
	\end{equation}
where $ G $ is the gravitational constant, $ c $ is the speed of light, $ D $ is the distance to the source, $ Q $ is the gravitational quadrupole, and $ \epsilon $ is the degree of asymmetry of the cocoon, which depends on the viewing angle $ \theta_{\rm obs} $ (see Fig.~\ref{fig:sketch}).
    \begin{figure*}
	\centering
		\includegraphics[scale=0.16]{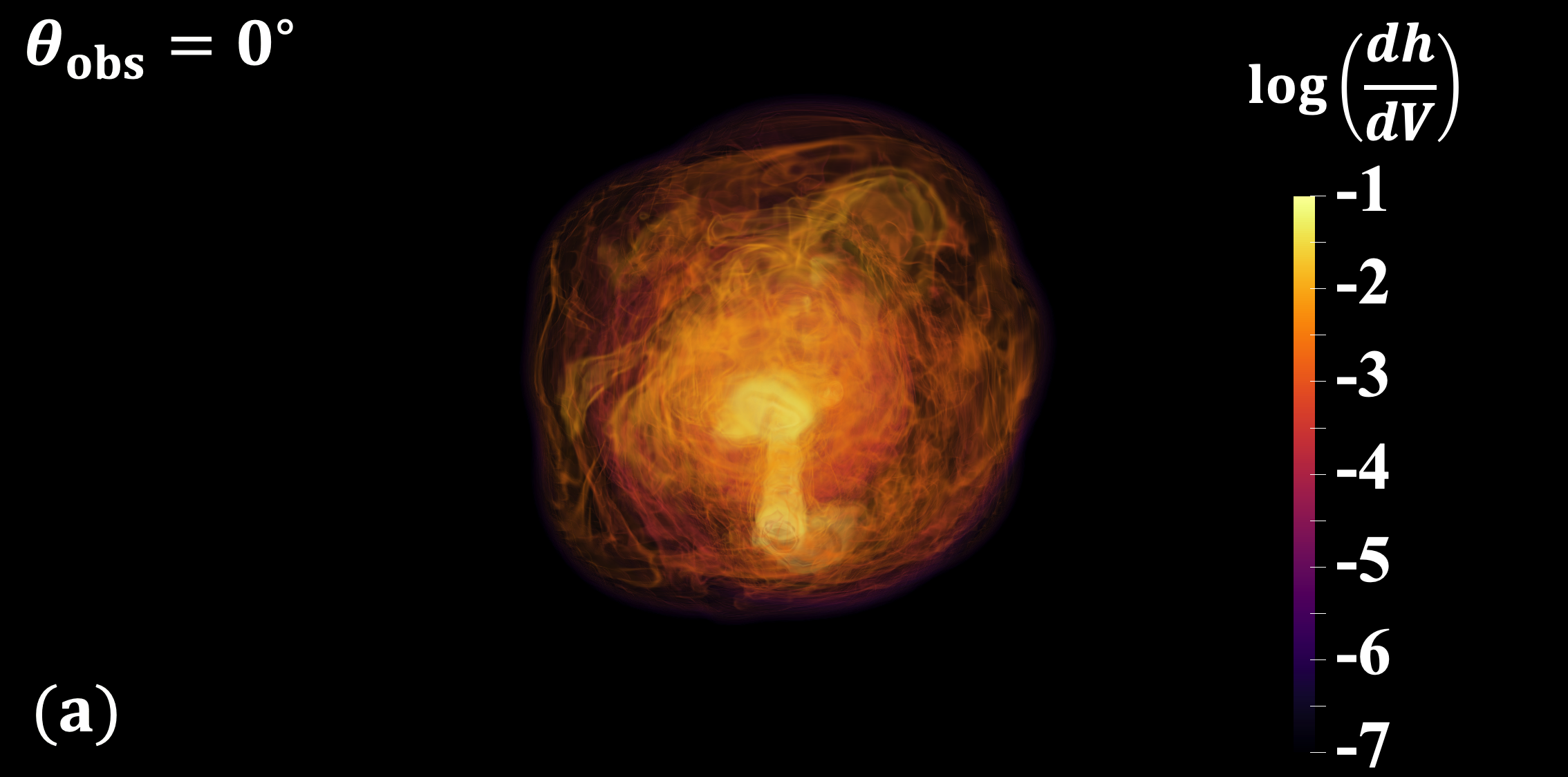}
		\includegraphics[scale=0.16]{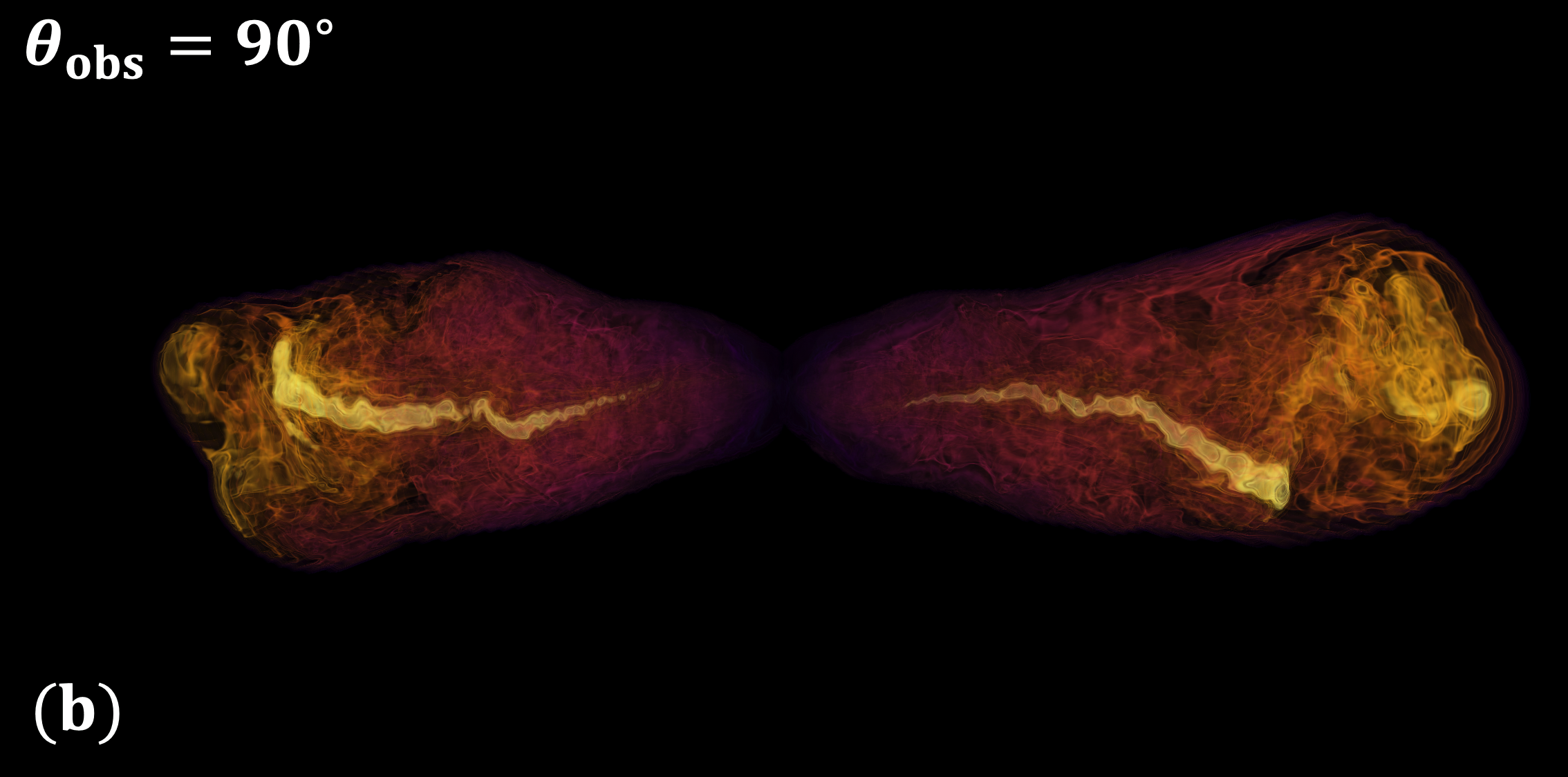}
		\caption{Three-dimensional (3D) rendering of the cocoon-powered GW strain density in c.g.s. units, as calculated in the lab frame without considering retarded time effects. The snapshots are when the cocoon breaks out from the star, 2.5 seconds after the BH formation -- the width of panel (b) is $ \sim 2R_\star = 8\times 10^{10} $ cm. At $ \theta_{\rm obs} = 0^\circ $ the cocoon projection appears nearly axisymmetric and results in a weak on-axis signal (a), whereas at larger observer angles the asymmetric shape of the cocoon leads to strong GW emission (b).
		}
		\label{fig:3Dmaps}
	\end{figure*} 
The quadrupole estimate does not consider phase cancellation between different GW emitting regions, as it assumes that the cocoon is smaller than the GW wavelength. Due to the invalidity of this assumption, we utilize the approach of applying the retarded Green's function to post-process the simulation data (see Appendix \ref{sec:gw_calculation}). Fig.~\ref{fig:3Dmaps} depicts the GW strain density of the hourglass-shaped cocoon upon breakout from the star (see the full animation and accompanying sonification in \url{https://oregottlieb.com/gw.html}). For on-axis observers (Fig.~\ref{fig:3Dmaps}a), the projected shape of the cocoon is close to circular ($ \epsilon \ll 1 $), similar to CCSNe, significantly suppressing the quadrupole moment. Off-axis observers (Fig.~\ref{fig:3Dmaps}b), on the other hand, will see an asymmetric outflow that propagates parallel to their field of view, with an order unity asymmetry $ \epsilon \lesssim 1 $, resulting in much stronger GW emission.

	\begin{figure*}
		\includegraphics[trim={0 5cm 0 0},scale=1.0]{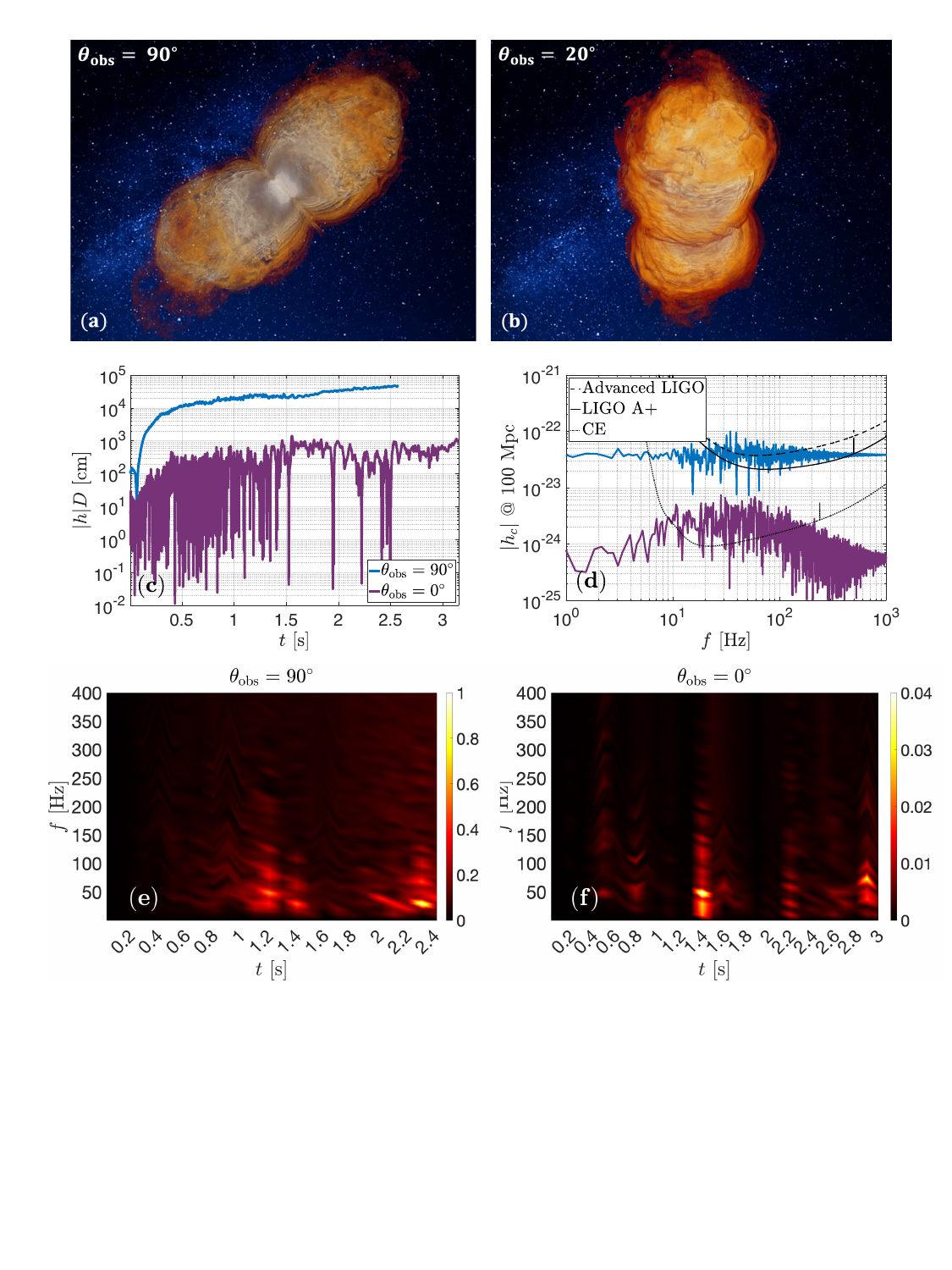}
		\caption{Top: 3D rendering of off-axis (a) and on-axis (b) projections of the cocoon mass density after breakout from the star (star is in the center shown in white), taken from \citet{Gottlieb2022b}.
        Middle:	On- and off-axis strain amplitudes of a typical LGRB event, shown in time (c) and frequency (d) domains.
		Bottom: Normalized strain amplitude (by maximum off-axis amplitude) spectrograms for a sliding window of 100~ms, shown for off-axis (e) and on-axis (f) observers.
		}
		\label{fig:cocoon}
	\end{figure*} 

Turbulent motions in the jet-cocoon outflow results in a stochastic GW burst with a broad spectrum of GW frequencies that is challenging to evaluate analytically. However, we can estimate the main features of the emerging GW spectrum as follows. 
The smallest length-scales of the cocoon emerge over the thickness of the shocked region $ \Delta r_{\rm sh} \simeq 10^{-2}R_{\rm sh}\Gamma^{-2} \lesssim 3\times 10^7 $ cm \citep{Nakar2012}, where $ \Gamma \simeq 2 $ is the jet head Lorentz factor inside the star, and $ R_{\rm sh} \lesssim 10^{10} $~cm is the radius of the shock. The cocoon elements are generated over characteristic timescale $ t_{\rm min} \simeq \Delta r_{\rm sh}/c_s \lesssim 5\times 10^{-4} $~s, where $ c_s \simeq c/\sqrt{3} $ is the relativistic sound speed, implying the highest GW frequency, $ f_{\rm max} \approx t_{\rm min}^{-1} \gtrsim 2000 $~Hz. The maximum timescale is the time that the jet energizes the cocoon, $ t_{\rm max} \gtrsim t_b \sim 10~$s, thereby setting the minimum GW frequency, $ f_{\rm min} \approx t_{\rm max}^{-1} \lesssim 0.1 $ Hz.
The cocoon energy is distributed quasi-uniformly in the logarithm of the proper-velocity \citep{Gottlieb2021a,Gottlieb2022b}, $ 10^{-2.5} \lesssim \Gamma\beta \lesssim 3 $. Thus, although various cocoon components evolve on disparate timescales, they carry comparable amounts of energy, and result in a rather flat GW characteristic strain between $ f_{\rm min} $ and $ f_{\rm max} $.
  	
The full GW calculation uses the retarded Green's function approach and reveals strong incoherent GW emission. The strain density maps in Fig.~\ref{fig:3Dmaps} demonstrate that the jets have the highest GW strain density. The wobbling nature of the jets, coupled with their substantial departure from axisymmetry, calls into question the validity of previous estimates which concluded that jets emit GWs only in sub-Hz frequencies, based on ideal, steady, and continuously axisymmetric top-hat jet models \citep[e.g.,][]{Leiderschneider2021}. However, when integrating the strain density over volume, we find that the observed GW emission is dominated by the freshly shocked material in the cocoon head, which occupies the majority of the outflow volume. We emphasize that the cocoon emission is not uniformly distributed throughout the cocoon. Had it been distributed uniformly, phase cancellations would have considerably attenuated the signal. Since most of the GW emission concentrates in smaller regions that are comparable in size to the characteristic GW wavelength, phase cancellations are partly suppressed, and a strong GW signal emerges. A comparison of the GW strain to the order of magnitude estimate in Eq.~\eqref{eq:h} shows that the calculated signal at $\theta_{\rm obs} = 90^\circ $ is weaker by a factor of $ \sim 3 $ than the maximum strain.

Figure~\ref{fig:cocoon}a,b shows a 3D mass density rendering of the cocoon after breakout from the star at $ t_b \approx 2.5 $~s, taken from \citet{Gottlieb2022b}. Similar to its pre-breakout shape in Fig.~\ref{fig:3Dmaps}, the cocoon is asymmetric when observed off-axis (Fig.~\ref{fig:cocoon}a), and near-axisymmetric when observed on-axis (Fig.~\ref{fig:cocoon}b). Fig.~\ref{fig:cocoon}d shows that the on-axis emission exhibits a distinct peak at $ 10~{\rm Hz} \lesssim f \lesssim 100~{\rm Hz} $, whereas the off-axis spectrum is rather flat with sporadic peaks in the same frequency range. The off-axis emission is stronger by 1-2 orders of magnitude than the on-axis emission, and by 2 orders of magnitude than the GW emission from CCSNe \citep{Abdikamalov2020}. Following Eq.~\eqref{eq:h}, we assume linear scaling of the strain amplitude with the outflow energy and calibrate the strain amplitude\footnote{We neglect secondary effects on the GW emission, such as the density profile of the star and turbulent mixing, that may affect the cocoon shape and its GW spectrum, and cause deviations from the linear scaling shown in Eq.~\eqref{eq:h}.},
    \begin{equation}\label{eq:calibrated}
	    h \approx 10^{-22}~\frac{40~{\rm Mpc}}{D}\frac{E}{10^{53}~{\rm erg}}~.
	\end{equation}
Approximating the GW emission to be isotropic, we find that the total GW energy is $ E_{\rm GW}\approx10^{50}$~erg, such that the GW conversion efficiency is $ E_{\rm GW} \sim 10^{-3}E $ (see Eq.~\eqref{eq:E_GW}). This implies that the GW efficiency is about two orders of magnitude higher than that in CCSNe, where $ E_{\rm GW} \approx 10^{-5}E_k $ \citep{Abdikamalov2020}, and $ E_k $ is the CCSN kinetic energy.

Throughout the entire simulation duration, the strain increases linearly with the energy injected into the system (note that Fig.~\ref{fig:cocoon}c is presented in a semi-logarithmic scale), consistent with Eq. \eqref{eq:h}.\footnote{We employ a full retarded calculation for off-axis observers, and neglect time-delay effects for on-axis observers, where the emission is weak (Appendix \ref{sec:gw_calculation}). Hence the difference in the integration times shown for off- and on-axis observers in Fig.~\ref{fig:cocoon}.}. The GW emission is expected to last until the jet engine shuts off (longer than our simulation), and the cocoon turbulent motions relax, hence the signal in Fig.~\ref{fig:cocoon} is expected to keep growing linearly with the jet activity time\footnote{GW travel time effects may slightly prolong the signal duration for observers close to the jet axis, see Appendix \ref{sec:gw_calculation}.}. The strain amplitude spectrogram (Fig.~\ref{fig:cocoon}e) shows that as the cocoon expands ($ t \lesssim 1 $~s in our simulation), the GW emission shifts only slightly toward lower frequencies, as the spectrum does not vary considerably between different regions in the cocoon, owing to mixing. Both on- and off-axis signals are qualitatively different from traditional GWs in CCSNe which have a peak frequency that rises over a much shorter GW emission timescale \citep[$ t \ll 1 $~s;][]{Kawahara2018,Powell2022}.

We calculate the matched-filter (MF) signal-to-noise ratio (SNR) for off-axis detection, and find $ \text{SNR}_{\text{MF}} = 2, 4, 28 $ at 100 Mpc in LVK O4, A+ and CE, respectively (see Eq.~\eqref{eq:SNR}). To assess the detectability, we conservatively employ a minimum SNR threshold of 20. This is motivated by the fact that a detection of a stochastic burst signal of this type will likely require the use of unmodeled search pipelines like coherent wave burst \citep{Klimenko2004,Klimenko2011,Klimenko2015} or STAMP \citep{Thrane2010ri,Macquet2021ttq}. Using this SNR threshold, we find that an outflow of $ E = 10^{53} $ erg could be detectable by A+ at about 20 Mpc. As we explain below, it is likely that some GRBs are associated with considerably more energetic outflows, for which the detection horizon is farther away. For on-axis observers, the GW detection horizon of such cocoons is smaller by 1-2 orders of magnitude, owing to the low degree of non-axisymmetry, which renders a direct on-axis detection unlikely.

	
\section{Detection rates}\label{sec:detectability}
        
    To estimate the number of detectable GW events in runs O4, A+ and CE, we use Eq.~\eqref{eq:calibrated}, connect $ E $ to the \emph{observed} jet energy via $ E = E_c + E_j \simeq 2E_j \approx \xi E_{j,\rm obs} $, and consider two types of jets with opening angle $ \theta_j \simeq 0.1^{+0.07}_{-0.03} $ rad \citep{Goldstein2016}: (i) a traditional axisymmetric jet, for which the observed jet energy is equal to the total jet energy, $ E_j = E_{j,{\rm obs}} $, i.e. $ \xi = 2 $, and conventional local LGRB rate $ {\cal{R}_{\rm GRB}}\sim 100 ~{\rm Gpc^{-3}}~{\rm yr}^{-1} $ \citep{Pescalli2015}; and (ii) a jet wobbling by $ \theta_w \simeq 0.2 $~rad as in our simulations, for which the local GRB rate is an order of magnitude lower \citep{Gottlieb2022b}, $ {\cal{R}_{\rm GRB}}\sim 10~{\rm Gpc^{-3}}~{\rm yr}^{-1} $. When a jet wobbles by $ \theta_w \simeq 2\theta_j $, it is observed on average only $ \left(\theta_j/\left[\theta_w+\theta_j\right]\right)^2 \simeq 10\% $ of the time, i.e. $ \xi \simeq 20 $. This implies that the outflow has more energy for a given $ \Egiso $, since most of its energy ($ \simeq 90\% $) is beamed away from our line of sight. Assuming the $ \gamma $-ray energy is $ 20\% $ of the total jet energy \citep{Frail2001}, the isotropic equivalent $ \gamma $-ray energy is $ \Egiso\simeq 0.2\times2\theta_j^{-2}E\xi^{-1} $. For example, in our simulation, the total outflow energy is $ E \simeq 10^{53} $ erg, the jet energy $ E_j \simeq 3\times 10^{52} $ erg and opening angle $ \theta_j \simeq 0.1 $ rad, so $ E_{j,\rm obs} \simeq 3\times 10^{51} $ erg, and $ \Egiso \simeq E $. Thus, although the simulated outflow energy may seem rather high, the observed $ \gamma $-ray jet energy is in fact fairly typical. We conclude that within the framework of our simulation, which features a wobbling jet, a typical GRB event might produce detectable GWs at 20 Mpc, making close-by SNe Ic-BL such as SN 2022xxf \citep{Kuncarayakti2023} viable candidates for GW detection.

    \begin{figure*}
    \centering
    	\includegraphics[scale=0.15]{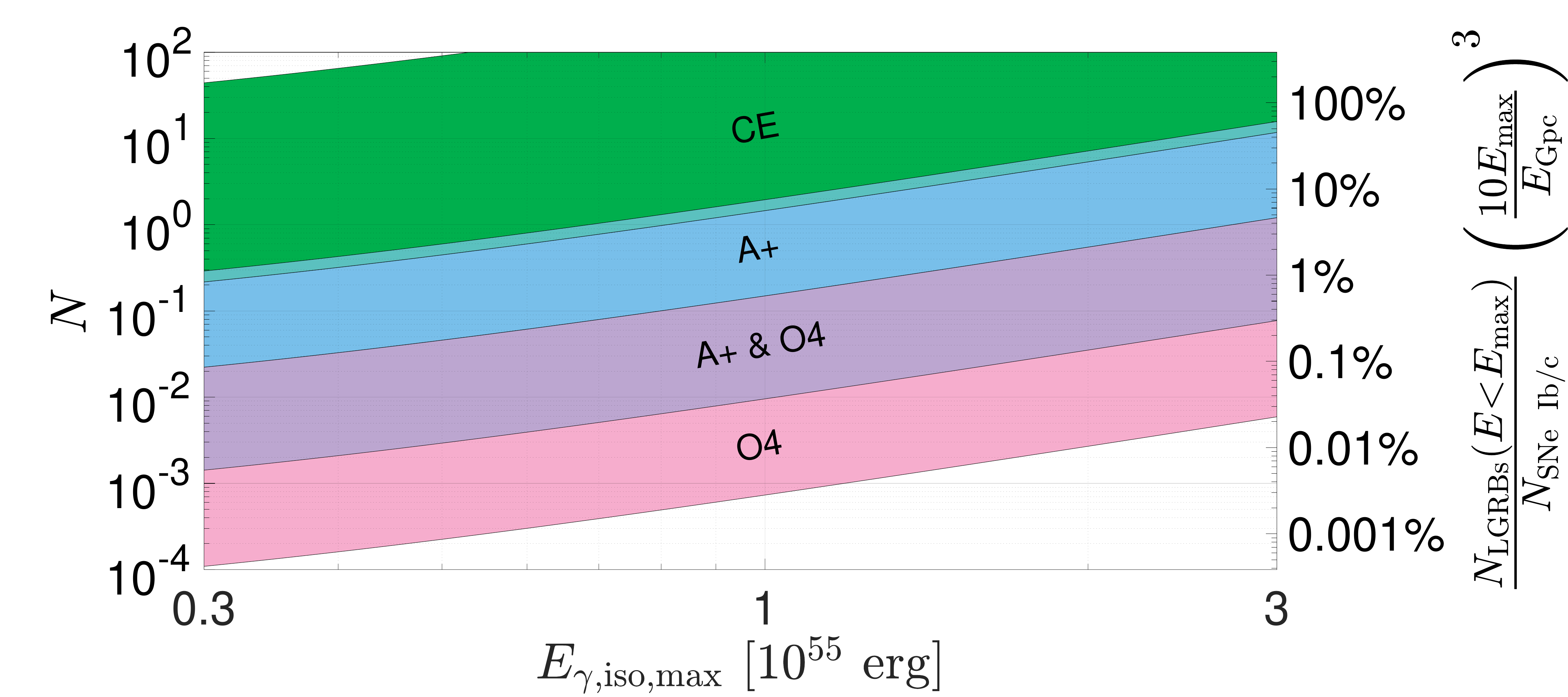}
    	\caption{The number of detectable GW events in CE (green) in 1 year, A+ (blue+purple) in 2 years, and O4 (pink+purple) in 1.5 years as a function of the maximum isotropic equivalent $ \gamma $-ray energy $ \Egisomax $ (the cutoff of the power-law fit to $ N_{\rm LGRBs}(E_{\gamma,{\rm iso}}) $ distribution), assuming MF SNR values, minimum SNR for detection of 20. These rates are calculated self-consistently with our simulation for wobbling jets. The range in the number of detectable events reflects the uncertainty in the jet opening angle. The detection rate of outflows with $ E < E_{\rm max} $ also indicates the abundance of LGRB events among SNe Ib/c (right vertical axis labels).}
    	\label{fig:detectability}
    \end{figure*}

    The {\it detected} LGRB population can be modeled by a log-normal distribution, whereas the {\it total number} of LGRBs can be fit by a power-law that coincides with the log-normal distribution tail at $ \Egiso > \Egisomin \approx 5\times 10^{53}~{\rm erg} $ (figure 5 in \citealt{Butler2010}). Thus, the tail of the distribution, which dominates the GW-detectable events can be constrained observationally. Integration over the distribution shows that LGRBs at this range comprise about half of the {\it detected} GRB population. The fit of the log-normal distribution tail to a power-law yields $ dN_{\rm LGRBs}/dE_{\gamma,{\rm iso}} = A \Egiso^{-2} $ \citep{Lan2022}, where the normalization $ A $ is set by the requirement that
    \begin{equation}
        \int_{\Egisomin}^{\Egisomax} A \Egiso^{-2}dE_{\gamma,{\rm iso}} = 0.5{\cal{R}_{\rm GRB}}~,
    \end{equation}
    thus
    \begin{equation}
        A = \frac{\cal{R}_{\rm GRB}}{2\left(\frac{1}{\Egisomin}-\frac{1}{\Egisomax}\right)}~.
    \end{equation}
    As the number of detectable events in a given volume grows as $ E_{\gamma,{\rm iso}}^3 $, the total number of detectable GW events scales as $ N \propto N_{\rm LGRBs}/dE_{\gamma,{\rm iso}}\Egiso^3dE_{\gamma,{\rm iso}} \propto \Egiso^{2} $ (Eq.~\eqref{eq:events}). Namely, the most energetic LGRBs dominate the detectable GW events, and the $ \Egiso $ power-law cutoff, $ \Egisomax $, sets the expected number of detectable events. The recent detection of GRB 221009A -- the brightest GRB of all time suggests that $ \Egisomax \approx 10^{55} $~erg \citep[e.g.,][]{Burns2023,OConnor2023}.

    For estimating the detectability in upcoming LVK observing runs, we assume the following: Linear scaling between the strain amplitude and the outflow energy (Eq.~\eqref{eq:calibrated}); $ \text{SNR}_{\text{MF}} = 2, 4, 28 $ at 100 Mpc for O4, A+ and CE runs, respectively; required SNR for detection of 20; isotropic GW emission to be similar to off-axis; and we consider only one detector for simplicity. We find that the number of detectable GW events per year is
    \begin{equation}\label{eq:events}
        N_{\rm yr} = \int_{\Egisomin}^{\Egisomax}\left(\frac{5\Egiso\frac{\theta_j^2}{2}\xi}{E_{\rm Gpc}}\right)^3A \Egiso^{-2}dE_{\gamma,{\rm iso}}~,
    \end{equation}
    where the parentheses reflects the fraction of detectable GW events in the energy band $ dE_{\gamma,{\rm iso}} $ within a given volume. The cubic dependency comes from the linear scaling of the GW strain amplitude with the energy, and inverse scaling with the distance. The numerator is the outflow energy, $ E = 2.5\Egiso\theta_j^2\xi $, and the denominator, $ E_{\rm Gpc} = 0.7, 5, 10\times 10^{54}~{\rm erg~Gpc^{-1}} $, is the energy required to reach SNR=20 at $ 1 $ Gpc in CE, A+ and O4 runs, respectively.
    
    The detection rate of the quasi-isotropic GW signal can also be used to estimate the abundance of LGRBs among their progenitors - SNe of type Ib/c, whose rate is $ {\cal{R}_{\rm SNe~Ib/c}} \approx 2.6\times 10^4~{\rm Gpc^{-3}~yr^{-1}} $ \citep{Li2011}. If the number of detected events with $ E < E_{\rm max} $ in the LVK observing run is $ N(E<E_{\rm max}) $, then the fraction of SNe Ib/c progenitors that power LGRBs with $ E < E_{\rm max} $ can be estimated as
    \begin{equation}\label{eq:SN}
         \frac{N_{\rm LGRBs}(E<E_{\rm max})}{N_{\rm SNe~Ib/c}} \approx \frac{N_{\rm yr}(E<E_{\rm max})}{\frac{{\cal{R}_{\rm SNe~Ib/c}}}{{\rm Gpc^{-3}~yr^{-1}}}\left(\frac{E_{\rm max}}{E_{\rm Gpc}}\right)^3}~.
    \end{equation}

The detection rate scales as $ N \sim \xi^3{\cal{R}_{\rm GRB}} $ (Eq.~\eqref{eq:events}), implying that the rate for axisymmetric jets is lower by two orders of magnitude. Therefore, our calculations assume that the majority of jets exhibit wobbling behavior, as observed in all first-principles collapsar simulations \citep{Gottlieb2022a, Gottlieb2022b}. Fig.~\ref{fig:detectability} displays the number of detectable GW events expected in upcoming LVK observing runs \citep{LIGOruns}, considering wobbling jets. If all jets are axisymmetric, these rates would be lower by two orders of magnitude. The uncertainty in the opening angle of the jet translates into a large uncertainty in the estimated number of detectable events, owing to its quadratic dependence on $ \theta_j $. The green range, which represents the number of detectable events in CE in one year, suggest that a GW detection is likely by CE. The blue and purple ranges imply a possible detection during the LVK A+ run, whereas in O4 run (pink and purple ranges) a detection is unlikely.
We emphasize that in addition to the large uncertainties in $ \theta_j $, the distribution of the most energetic GRBs and the prompt/afterglow energy ratio \citep{Beniamini2016} are poorly constrained, and introduce an even larger range of uncertainty into the expected number of detectable GW events\footnote{Additionally, in our estimate we ignore jets that are choked in the stellar envelope and do not generate a GRB. If such phenomenon is common among massive stars \citep{Gottlieb2022a}, then the predicted GW detection rate is increased significantly (see \S\ref{sec:conclusion}).}. Interestingly, the detection rate of these quasi-isotropic GWs with energies $ E < E_{\rm max} $ constrains the fraction of SN Ib/c progenitors that power LGRBs with $ E < E_{\rm max} $, as shown on the right vertical axis in Fig.~\ref{fig:detectability}. In order to optimize the GW detection, we need: (i) unmodeled searches, the characterization of which would require GW signatures from different jetted explosions to understand the diversity of signals (this will be addressed in future work); (ii) good localization of the source, which can be achieved in the likely case that the GW signal is accompanied by another detectable messenger signal, as we discuss below.

\section{Multimessenger events}\label{sec:multimessenger}
   
Although joint LGRB-GW detection is unlikely due to the weak on-axis GW emission, the GW signal is likely accompanied by a wide range of electromagnetic counterparts powered by the SN explosion and the cocoon: shock breakout in $ \gamma $- and X-rays (seconds to minutes), cooling emission and radioactive decay in UV/optical/IR (days to months), and broadband synchrotron (afterglow) emission (days to years) \citep{Lazzati2017,Nakar2017}. For close-by CCSNe, neutrino detection is also possible, and will place even stronger constraints on the expected time of the GW signal. Thus, LGRBs are rich multi-messenger sources. The earliest radiative signal emerges when the cocoon or SN shock wave breaks out from the star, producing a nearly coincident electromagnetic counterpart to the GWs. However, the shock breakout signal originates in a thin layer, and primarily depends on the breakout shell velocity and structure of the progenitor star, rather than the total explosion energy \citep{Nakar2012,Gottlieb2018b}. Thus, although under favorable conditions of viewing angle and progenitor structure it is possible to detect a shock breakout in $ \gamma $-rays, those signals will typically go unnoticed.

After releasing the shock breakout emission, the stellar shells and the cocoon expand adiabatically and give rise to an optical cooling signal. SNe Ib/c cooling emission lasts weeks and peaks at an absolute magnitude $ M_{\rm AB} \approx -18 $ \citep{Drout2011}, whereas cocoons with energies of $ E_c \sim 10^{52} $~erg that emit strong GWs, will power even brighter ($ M_{\rm AB} \sim -19 $) quasi-isotropic ($ \theta_{\rm obs} \lesssim 1.0 $ rad) UV/optical cooling emission on timescales of days \citep{Nakar2017,Gottlieb2018a}. The optical emission of the cocoons and SNe is sufficiently bright to be detected at all relevant distances of a few hundred Mpc by Zwicky Transient Facility (ZTF) \citep{Graham2019,Masci2019}. However, the cocoon cooling emission lasts a few hours, challenging the detection of such signals. The prospect of future telescopes such as the Rubin Observatory \citep{LSST2009} and ULTRASAT \citep{Ben-Ami2022, Shvartzvald2023} holds great promise in significantly enhancing the rate of such detections. These advanced observatories will provide valuable insights into the energetics and abundance of these events. Furthermore, cocoons with $ E_c \gtrsim 10^{52}~{\rm erg~s^{-1}} $ will explode the entire massive star and may power superluminous supernovae (SLSNe) with $ M_{\rm AB} \lesssim -21 $ over $ \sim $ months timescale \citep{Gal-yam2012}.
    
Finally, the interaction of the cocoon with the circumstellar medium will produce a detectable broadband afterglow, assuming typical ambient densities around collapsars and standard equipartition parameters \citep{Gottlieb2019}. The timescale over which the afterglow emission emerges varies from days to years, as it depends on the specific parameters of the system and the observer's viewing angle, thereby posing a challenge to its detection. However, the early cooling signal, which can be detected by a rapid search will enable an early localization of the event and a targeted search for the later multi-band afterglow, which will potentially alleviate the afterglow detection difficulty.

\section{Conclusions}\label{sec:conclusion}
        
In this {\it Letter}, we propose and investigate a new non-inspiral GW source originating in the cocoon that is generated during the jet propagation in a dense medium. The cocoon is a promising GW source for the following reasons: (i) it contains at least as much energy as the jet, and likely even more considering the jet wobbling motion; (ii) when observed off-axis, its projected shape is highly asymmetric, giving rise to a substantial gravitational quadrupole moment; (iii) it evolves over considerably shorter timescale than the jet, such that its GW emission peaks at $ 10~{\rm Hz} \lesssim f \lesssim 100~{\rm Hz} $. However, we note that the presence of an unstable and wobbly jet may also lead to GW production within the frequency bands detectable by LVK. This raises concerns and casts doubt on the validity of previous numerical and analytic calculations of GWs originating from continuously launched, steady, and axisymmetric jets. We find that collapsar cocoons are the brightest stochastic GW sources in LVK frequency bands known to date, and could be detectable out to dozens of Mpc in LVK A+ run.

The jet structure plays a crucial role in determining the detection rates. If the jet wobbles, as indicated by state-of-the-art first-principles simulations, the cocoon energy may be as high as $ E_c \gtrsim 10^{52} $ erg. We note that if the jet wobbles, the inferred $ \gamma $-ray efficiency is lower than its real value due to emission wasted in the direction out of the line of sight, which would also result in less energy available for the afterglow emission. However, the confirmation of such high energies still requires observational evidence, which could be provided by a large sample of optical emission detections from collapsar cocoons. Thus, the upcoming telescopes, such as the Rubin Observatory and ULTRASAT, hold the potential to shed light on the energy properties of these cocoons. If indeed jets wobble, the detection rate of their cocoons is expected to be about two orders of magnitude higher compared to the rates considered for traditional axisymmetric jets. For example, the estimated detection rate in the third-generation GW detectors is $ \sim 10 $ events per year for wobbling jets. For axisymmetric (wobbling) jets, the detection rate is only $ \sim 0.1 $ events per year in third-generation GW detectors (LVK run O4).

Another noteworthy jet-cocoon populations for production of detectable GWs are short GRB jets in binary neutron star (BNS) mergers, and jets that fail to break out (and deposit all their energy in the cocoon) either from the BNS ejecta \citep{Nagakura2014} or from the star \citep{Gottlieb2023}. Choked jets could be even more abundant than those that successfully break out, based on the GRB duration distribution \citep{Bromberg2012}. However, LGRB jets with short central engine activity are unlikely to drive the powerful cocoons required for producing a detectable GW signal, as their cocoon structure will be quasi-spherical due to early choking. Alternatively, if jets fail because of the progenitor structure, e.g. it has an extended envelope such as in SN Ib progenitors \citep{Margutti2014,Nakar2015}, their cocoon could be very powerful, and even give rise to energetic explosions and electromagnetic signatures such as fast blue optical transients \citep{Gottlieb2022c}. In this case, the GW detection rate can be significantly higher. The rate of short GRBs is $ 3-30 $ (depending on whether jets wobble) higher than that of LGRBs \citep{Abbott2021a}, but they are also about an order of magnitude less energetic \citep[e.g.,][]{Shahmoradi2015}. Because the detectable GW events are dominated by the most energetic GRBs, it is unlikely that LVK run A+ will detect cocoons that accompany short GRB jets in binary mergers, but the CE might be able to detect a few.

Future calculations of the zoo of cocoon-powered GWs for different LGRB progenitors and other cataclysmic events will enable the characterization of amplitude/spectrum based on the source properties, and will be addressed in follow-up studies. This will enable more efficient searches for these GWs, enhance the wealth of information regarding the physical properties of the sources to be extracted from the GW signals, and aid follow-up searches for late electromagnetic counterparts (e.g., afterglow) of these multi-messenger events, ultimately providing a better understanding of the relation between CCSNe and LGRBs.

\begin{acknowledgements}

We thank the anonymous referee, Yuri Levin, Kenta Hotokezaka, David Radice, Fabio De Colle, Brian Metzger and Ilya Mandel for valuable comments.
HN thanks Hirotada Okawa for useful comments and discussions.
OG is supported by a CIERA Postdoctoral Fellowship.
OG and AT acknowledge support by Fermi Cycle 14 Guest Investigator program 80NSSC22K0031.
AT was supported by NSF grants
AST-2107839, 
AST-1815304, 
AST-1911080, 
AST-2206471, 
OAC-2031997, 
and NASA grant 80NSSC18K0565. 
PN gratefully acknowledges support at the Black Hole Initiative (BHI) at Harvard as an external PI with grants from the Gordon and Betty Moore Foundation and the John Templeton Foundation. ER-R thanks the Heising-Simons Foundation and the NSF (AST-1911206, AST-1852393, and AST-1615881) for support.
This research used resources of the Oak Ridge Leadership Computing Facility, which is a DOE Office of Science User Facility supported under Contract DE-AC05-00OR22725. An award of computer time was provided by the ASCR Leadership Computing Challenge (ALCC), Innovative and Novel Computational Impact on Theory and Experiment (INCITE), and OLCF Director's Discretionary Allocation  programs under award PHY129. This research used resources of the National Energy Research Scientific Computing Center, a DOE Office of Science User Facility supported by the Office of Science of the U.S. Department of Energy under Contract No. DE-AC02-05CH11231 using NERSC award ALCC-ERCAP0022634.

	\end{acknowledgements}

\section*{Data Availability}
	
	The data underlying this article will be shared upon reasonable request to the corresponding author.
	All numerical data used in this paper will be shared upon request to the corresponding author.

\bibliography{refs_apjl}

\appendix

\section{Gravitational-wave calculation}\label{sec:gw_calculation}
 
    When a fluid particle accelerates, it induces a perturbation in the metric, which results in emission of GW radiation. The GW strain at the observer frame calculated using the retarded Green's function\footnote{This approximation is valid in flat space-time. Since in the innermost $ \sim 100 r_g $ ($ r_g $ is the BH gravitational radius) there is a negligible amount of energy (the cocoon is many thousands of $ r_g $, e.g. Fig.~\ref{fig:3Dmaps}), it can be applied to the calculation of GWs from cocoons.}
	\begin{equation}\label{eq:h2}
	    h_{\mu\nu}(t,x) \approx \frac{4G}{Dc^4}\int\frac{T_{\mu\nu}\left(x',t-\lvert x-x'\rvert/c\right)}{\lvert x-x'\rvert}dx'dy'dz'~,
	\end{equation}
    where $ T_{\mu\nu} $ is the covariant stress-energy tensor, and the integration is over the volume of the system. For the numerical calculation, we evaluate the stress-energy tensor for each cell in the simulation. Subsequently, we divide the simulation grid by slicing it along surfaces parallel to the $ \hat{y} $-axis, creating a series of $ \hat{x}-\hat{z} $ planes with a width of $ 10^3~{\rm km} $, where the $ \hat{z} $-axis represents the axis of rotation. For each plane, we integrate the total $ T_{\mu\nu} $ at every laboratory time. To account for time-delay, we calculate the retarded contribution of each plane by determining the retarded times as $ t_{\rm obs} = t - \lvert x - x'\rvert/c $. This is done with respect to an observer at infinity along the $ \hat{y} $-axis. For the faint on-axis emission, we neglect time-delay effects and assume that the observer time coincides with the laboratory time.
	
    For a single GW with angular frequency $ \omega $, that -- without the loss of generality -- propagates along the $ \hat{x} $-axis, a plane wave solution has two independent components in the transverse-traceless (TT) gauge metric
    \begin{equation}\label{eq:hij}
        h_{\mu\nu}(t,x)=\left(
        \begin{matrix}
        0 & 0 & 0 & 0 \\ 0 & 0 & 0 & 0 \\ 0 & 0 & h_{+} & h_{\times} \\ 0 & 0 & h_{\times} & -h_{+}
        \end{matrix}
        \right){\rm cos}\left[\omega\left(t-\frac{x}{c}\right)\right]~,
    \end{equation}
    where
    \begin{equation}\label{eq:polarizations}
    \begin{split}
        h_+ = h_{yy}-h_{zz}~, \\
        h_\times = 2h_{yz}~
    \end{split}
    \end{equation}
    are the two independent polarizations of the plane wave moving along the $ \hat{x} $-axis.
    
    The GW detector response function to the planar GW is a linear combination of the two polarizations \citep{Jaranowski1994} and is given by
    \begin{equation}\label{eq:superposition}
        h(t) = F_+h_+(t) + F_\times h_\times(t)~,
    \end{equation}
    where $ F_+, F_\times $ are functions of the antenna-pattern of the detector that depend on the sensitivity to each polarization. For an order of magnitude estimate, the GW strain can be approximated as
    \begin{equation}\label{eq:ht}
    	h(t) \approx \sqrt{h_+^2(t)+h_\times^2(t)}~,
    \end{equation}
    and similarly the Fourier transform of the interferometer response to the dimensionless GW strain
    \begin{equation}\label{eq:fourier}
        \tilde{h}(f) = \sqrt{\tilde{h}_+^2(f)+\tilde{h}_\times^2(f)}~.
    \end{equation}
    Finally, the GW energy in the isotropic approximation is \citep{Chatziioannou2017}
    \begin{equation}\label{eq:E_GW}
        {E}_{\rm GW} = \frac{c^3}{G}\pi^2D^2\int_{-\infty}^{\infty}\left(\lvert \tilde{h}_\times(f)\rvert^2+\lvert \tilde{h}_+(f)\rvert^2\right)f^2df~.
    \end{equation}
    To characterize the sensitivity for detection at a given frequency, one defines the characteristic strain as \citep{Moore2015}
    \begin{equation}\label{eq:hc}
        h_c(f) = 2f\lvert \tilde{h}(f)\rvert~.
    \end{equation}
    
We estimate the MF SNR as the inner product of the strain, whitened by power-spectral density of the noise $ S(f) $ \citep{Moore2015}
\begin{equation}\label{eq:SNR}
    \text{SNR}_{\text{MF}} = \sqrt{\int \frac{4 |\tilde{h}(f)|^2}{S(f)}}df~.
\end{equation}
We use the \textsc{PyCBC} package~ \citep{Usman2016,Nitz2021} for the noise power-spectral densities of advanced LIGO, A+ and CE detectors. 

	
For the numerical simulation, we follow \citet{Gottlieb2022b} who performed high resolution 3D GRMHD simulations of a collapsing star with strong magnetic fields (maximum B-field in the stellar core at the time of the BH collapse is $ B \approx 10^{12.5} $ G). We use the code \textsc{h-amr} \citep{Liska2022} to rerun one of their simulations (with maximum initial magnetization of $ \sigma_0 = 15 $), but double the mass scale of the simulation such that the stellar mass is $ M_\star = 28 \msun$, the stellar radius is $ R_\star = 4\times 10^{10} $~cm. The Kerr BH mass is $ \mbh = 4.2\msun $ and its dimensionless spin is $ a = 0.8 $. We follow the outflow evolution until the jet head breaks out from the collapsing star and reaches $ \simeq 1.5R_\star $ at $ t \simeq 3.2 $~s with asymptotic Lorentz factor $ \sim 5 $. We output data files every $ \Delta t = r_g/c \approx 0.02 $ ms. We verify that the strain does not change when increasing the spatial resolution in each dimension by a factor of 1.5. The integration to $ t \gtrsim 3~{\rm s} \approx 150,000~r_g/c $ implies that we post-process 150,000 files with more than a petabyte of data.

\end{document}